# Development of A Fully Data-Driven Artificial Intelligence and Deep Learning for URLLC Application in 6G Wireless Systems: A Survey

Adeeb Salh[1], Lukman Audah[2], Qazwan Abdullah[3], Abdullah Noorsaliza[4,] Nor Shahida Mohd Shah[5], Jameel Mukred[6], Shipun Hamzah[7]

**Abstract.** The full future of the sixth generation (6G) will develop a fully data-driven that provide terabit rate per second, and adopt an average of 1000+ massive number of connections per person in 10 years (2030 virtually instantaneously). Data-driven for ultra-reliable and low latency communication (URLLC) is a new service paradigm provided by a new application of future 6G wireless communication and network architecture, involving 100+ Gbps data rates with one-millisecond latency. The key constraint is the amount of computing power available to spread massive data and well-designed artificial neural networks (ANNs). Artificial Intelligence (AI) provides a new technique to design wireless networks by apply learning, predicting, and make decisions to manage the stream of big data training individuals, which provides more the capacity to transform that expert learning to develop the performance of wireless networks. We study the developing technologies that will be the driving force are artificial intelligence, URLLC in 6G communication systems to guarantee low latency. This paper aims to discuss the efficiency of the developing network and alleviate the great challenge for application scenarios and study Holographic radio, enhanced wireless channel coding, enormous Internet of Things (IoT) integration, and haptic communication for virtual and augmented reality provide new services on the 6G network. Furthermore, improving a multi-level architecture (MLA) for URLLC in deep Learning (DL) allows for data-driven AI and 6G networks for device intelligence, as well as allowing innovations based on effective learning capabilities. These difficulties must be solved in order to meet the needs of future smart networks. Furthermore, this research categorizes various unexplored research gaps between machine learning (ML) and 6G.

## INTRODUCTION

Wireless networks of the fifth generation (5G) have just lately began to be used in fully intelligent networks. Developing the role of intelligence, the Internet of Everything (IoE), and relationships in the edge intelligence, device intelligence at the mobile user, and edge infrastructure components will be driving forces in the 6G generation of wireless communications architecture. [1][2]. The sixth-generation (6G) employs high-frequency channels to transmit large amounts of data. As demonstrated in Figure 1, THz's artificial intelligence (AI) technology and huge bandwidth enable high-rate, reliable low latency by enabling a large coverage that enables connected intelligence. AI creates and solves the most important technologies in the transmission network, such as route management, topology control, security, and secrecy. In addition, maximize the high video transmissions in the long term and intelligent schedule for packet transmission in 6G depend on enabling learning with IoT. Furthermore, the terahertz (THz) enables high-data rates that are driving the optimization of 6G wireless networks, including ML and deep learning (DL). Although various components of 5G wireless networks have just begun to evolve, the goal of 6G will be to create a fully immersive user experience that functions as a fully intelligent system and gives everything. Microwave transmissions over the sub-6 GHz band were used in the previous generation of 5G [4-12], which are impossible to achieve in millimeter-wave frequencies, where the system bandwidth barely exceeds 1 GHz. Supporting high rates and excellent dependability necessitates the utilization of THz's large bandwidth. The demand for dependability, such as the likelihood of packet loss $10^{-5} \sim 10^{-7}$. The length of time that the training sample has been used It will not be desirable if a device's packet arrival rate is low, necessitating the use of DL in URLLC to send more than packets in a real-time network in order to optimize the multi-level architecture (MLA) for data-driven applications [13-19].

Advanced ML techniques, ML optimized edge computing networks with the advancement of URLLC, and distributed artificial intelligence are the major tools for achieving AI. Moreover, holographic communications and high precision manufacturing improved depends on a huge artificial brain that providing practical zero-latency services, indefinite storage, and massive cognition capabilities. The 5G are not supported the high demand of data rates to Tbps, because the holographic communications need to utilize view multiple cameras with providing an actual experience. In 6G, entirely data-driven approaches are emerging, utilizing a deep learning and integrated artificial neural network that used correct mathematical models for the original network in all phases of network creation. Also improving by reducing the amount of real-time data that must be learned and estimated in order to use data-driven strategies.

In this paper, we explain the overarching goal for URLLC in 6G, and we cover six important technologies that will give a survey on various ML methodologies realistic to communication, based on AI-enabled intelligent 6G networks. The goal of this analysis is to present a thorough picture of DL systems:
- To highlight critical URLLC needs and achieve the demanding condition of 6G wireless network visions under a new enabling AI in URLLC that the wireless research community should pay special attention to. Its main contribution is to apply AI to advanced ML approaches by learning, forecasting, and making judgments to control the flow of huge data training individuals, as well as to increase a MLA for data-driven DL.
- Major difficulties that will need to be addressed in the coming years, relying on the development of DL to give high computing efficiency for the construction of efficient artificial intelligence. To pique future research interests in the proposed 6G network, which is built on efficient learning capabilities and a MLA that enhances the 6G network.

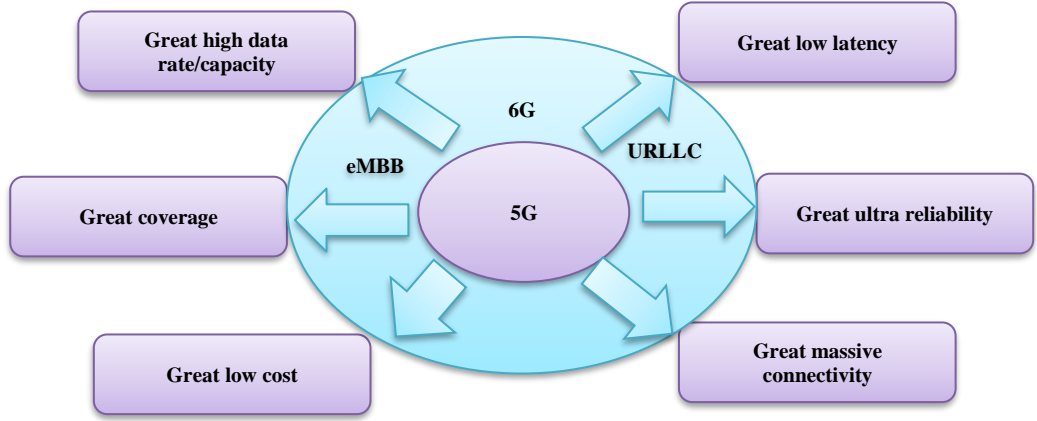

**FIGURE 1**: Requirements of 6G networks for URLLC.

## NEW SERVICES OF 6G

To achieve the URLLC, develop a traffic-flow prediction in 6G based on the short-term correlation prediction for the intelligent IoT and control architecture. Manage more traffic flow by delivering packets in a real-time network via a wireless channel based on efficient, accurate data-driven forecasts, which can increase the MLA for data-driven decisions [13].

**TABLE 1:** Present a Summary of the Developing Wireless Technology [6], [9].

| Evolving wireless technologies | 2020 5G | 2030 6G |
|---|---|---|
| Frequency bandwidth (bandwidth) | 3-300 GHz | 1-10THz |
| Bandwidth | 0.25-1 GHz | Up to 3THz |
| Data rate | Up to 20 GHz | Greater than 1 THz |
| Mobility | Up to 500km/h | Up to 1000km/h |
| Latency | 10ms | Less than 1ms |

The new 6G services are organized around four new nature-inspired technologies: holographic radio, enhanced wireless channel coding, vast IoT integrated and haptic communication, and Tactile Internet. Concentrate on critical technologies that are too small for 5G and require new KPI drivers for 6G wireless networks, as shown in Table2.

## Holographic Radio

The ability to communicate in holograms is a feature that draws people to 6G wireless networks. The URLLC is a breakthrough in delivering holographic videos that requires a large amount of spectrum bandwidth, which is available in the THz bands of 6G. Furthermore, holography calling comprises of holographic massive input massive output and 3-D spectral holography, both of which can work in the 6G due to the usage of vast intelligent surfaces and full closed-loop control of the entire physical space during spatial-spectral holography. To facilitate holographic and high-precision communications, two drivers for 6G networks for mobile Internet and IoE are being used.

**TABLE 2:** KPI evaluation between 5G and 6G wireless communication [6], [10],[19].

| Key Performance Indicators (KPI) | 5G | 6G |
|---|---|---|
| Traffic capacity | 1Gbps | 100Gbps |
| Highest Data rate in Downlink | 20Gbps | >1Tbps |
| Highest Data rate in uplink | 10Gpbs | >1Tbps |
| User experience | 50bps | 10Gbps |
| End to end latency | 1ms | <1ms |
| Mobility provision | Up to 500km/hr | Up to 1000km/hr |
| Processing Delay | 50ns | 10ns |
| Spectral and Energy Efficiency Gains | 10x in bps/Hz/m$^2$ | 1000x in bps/Hz/m$^3$ |
| Frequency Range | 3-300GHz | 10THz |
| Reliability | 99.999% | 99.99999% |
| Joining density | $10^6$ devices/km$^2$ | $10^8$ devices/km$^2$ |
| Power consumption | Medium | Relatively low |
| Haptic Communication | Partial | Fully |
| Security | Medium | High |
| Artificial intelligence | Partial | Fully |

## Wireless Channel Modeling for 6G

At Mm-Waves and THz frequencies, new channel evaluation algorithms for targeted broadcasts will become an important component of the 6G communication infrastructure. The fundamentally predictive leveraging of the recent advancement in ML is required to support very accurate channel predictions in huge URLLC [20-25]. To ensure the execution of the high engagement, good immersion into a distance requires the usage of holographic communication. [22]. Exponentially increase the speed of big data transmission by allowing full-CSI; In this case, the modeling of the channel is opened when transmitting data between the omnidirectional and receiving antennas [23]. The proposed channel distribution is based on the DL methods of generative models, smart frames to show the geometric distribution of channel dimensions, and channel model optimization and reception analysis provide high speed when transmitting a signal based on the proposed channel distribution. [24]. In addition, train deep neural networks for transmission prediction and improve the beams and propagation channels provided by using time-varying channel state information via customizable intelligent antennas [25]. The AI used lengthy block-length codes for channel decoding, based on the usage of a tanner graph learned for deep neural networks in order to obtain correct channel state information, and developed the combination with traditional training data to achieve performance advantages [26].

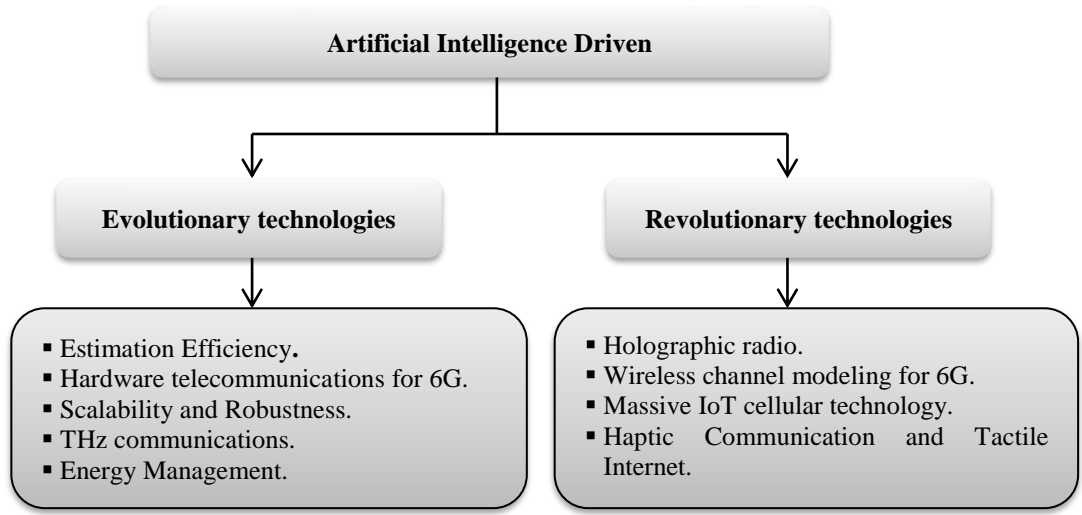

**FIGURE 2:** Essential characteristic to achieve the aforementioned 6G networks.

## Massive IoT Cellular Technology

IoT is an excellent processing device and its surroundings are sensitive to multiple interconnected devices, data storage, and data processing capabilities when working with artificial intelligence. Combining AI with IoT can provide better visibility and control over many Internet-connected devices. To meet the high demands in various applications, delay sensitivity, ubiquitous connectivity, big data analytics, high level of complexity of the device and wide range of power consumption on several mobile phones have been improved, using mass IoT in 6G mobile networks. Apply massive IoT in 6G cellular networks with intelligence transmission packet rate provide excellent coverage, high volume of low cost, and low cost of power consumption with high throughput. The high-dimensional data, convoluted decision productions achieve high-quality system performance without relying on perfect labeled data by applies the online training for reinforcement learning [27], [28].

## Haptic Communication and Tactile Internet

The amount of large data generated by wireless connections continues to grow, and it is often used when the latency is low enough to provide haptic communication for practical increased reality items in a variety of settings. In traditional multimedia services, URLLC's haptic communication can offer great quality video and audio traffic. The control of physical communication in close proximity to actual sights of environments during the tactile internet in real-time will be enabled by holographic communication of a virtual vision in 6G [28]. Furthermore, using the URLLC system in extended reality, the human brain is unable to discern between varied delays. The enhanced AI-enhanced computing proximity (Mobile Edge-Cloud) is proposed by offering personal knowledge to change the haptic trail to keep a person away from the loop (Tactile Internet) and to provide advance data traffic. [29], [34]. Because tactile Internet really serves a very important aspect of society, it needs to be very reliable and allow many devices to communicate with each other at the same time. It also needs support without delaying the flight to the end, otherwise, the touch user will experience "cyber-pain" and players and people will use poorly flying simulators. Tactile Internet can connect to traditional wired Internet, mobile Internet, and the Internet of Things, thus creating the Internet in entirely new dimensions and capabilities.

## ENABLES

This section discusses 6G wireless network visions in a URLLC with a different enabling AI. To meet the demanding conditions of the most recent service 6G, to conduct a survey on various ML approaches that are realistic for communication and networking, and to imagine solutions to enable AI in a future 6G wireless network. Figure 3 depicts the AI framework for wireless communication and mobility management, which is based on deep neural networks.

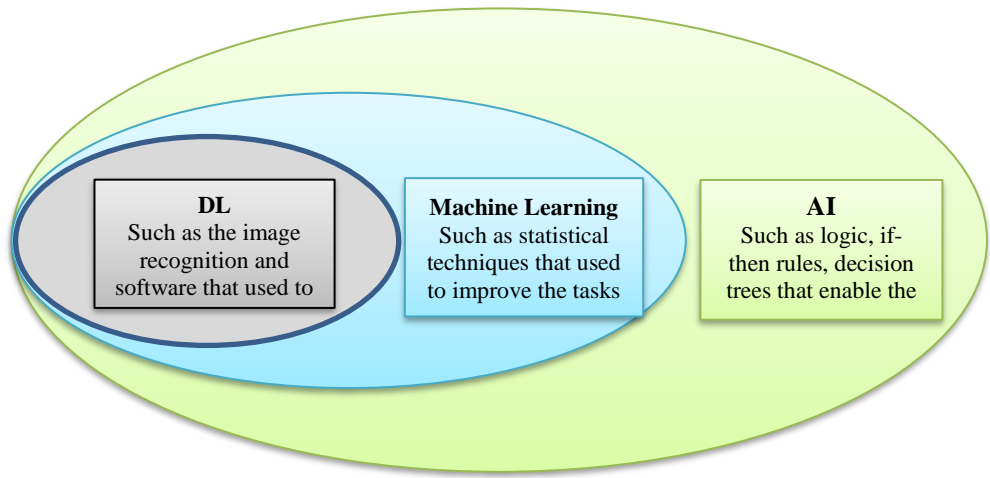

**FIGURE 3:** Artificial intelligence, ML, and DL relevance in 6G wireless communications.

### ML Enabled Intelligent 6G Networks

Machine learning and artificial intelligence (AI) are breakthrough technologies that develop system level explanations in 6G networks and the Internet of Things [30-39]. The flexibility of end-to-end delay and reliability requirements is determined by how the DL is used in the URLLC. By using data-driven ways through treating accessible data and increasingly learning, the ML algorithm is able to perform big data rates. In MLA, ML for artificial neural networks can establish and accomplish tasks such as $A: z \in Z \subseteq X^n \rightarrow b \in B \subseteq X^n$ where z is the input vector, b is the ML algorithm's output, A is the desired performance function, and by $z \in X^n$, $Z$ The set of z and b is represented as Z and B. Using the most effective activation function $a_{n,l}$ to $s_{n,l}$ includes fully linked layers, which are required for artificial neural networks, and select for the hidden layers' activation function [36]. The output of neuron n for layer l-th in the network is the result of the processing performed in each neuron n for layer l-th in the network. $z_l(n) = a_{n,l}(s_{n,l})$, where the intermediate term $s_{n,l} = \beta^T{}_{n,l} z_{l-1} + \rho_{n,l}$, The training process in supervised learning was improved by reducing loss in order to obtain the desired input-output.

The new issues in 6G wireless connectivity for mobile ML, which must be addressed by processing data locally and storing large amounts of data, rely on efficient distributed training to reduce computing complexity. The huge data was gathered through the investigation of ML strategies that are capable of evaluating and disciplinary approaches such as (supervised, unsupervised, and reinforcement learning) [38-44]. We introduce ML in the context of mobile and wireless communication networks in this section, which can be summarized as follows:

**Supervised Learning for URLLC:** Based on the considered training data accessibility and real-time, supervised learning has implemented the motivations of numerous circumstances in wireless networks. Furthermore, supervised learning used Ultra-Wide Band in real-time to find problems based on multiclass hypothesis predictions [38]. Owing to the extensive training of prediction mistakes when utilizing supervised DL in URLLC, the missed detection probability happened due to the large interference and communication collision caused at the beginning of each prediction:

**Unsupervised Learning:** By utilizing the experienced data training based on the labeled training data set and accompanied by the conforming desired output, it is possible to address learning non-deterministic problems and make real-time decisions. URLLC [38], [44] Learn the latent function with unsupervised DL to optimize train deep neural networks (DNN), and also to achieve high availability with simple loss of bandwidth based on QoS requirements. Users send data packets to rely on traditional statistics, such as a detailed probability analysis in the physical layer protocol, to verify the authenticity of the packet from the strength of the transmitted signal or the value of the channel [45].

**Deep deep-RL:** In smart cities, it is designed to provide a modeless URLC and highly dynamic transmission, as well as to reserve a reliable wireless connection for overhead networks based on resource allocation in 6G networks. A detailed review of the deep RL and neural networks can be found [53]. In recent years, advanced games, robotics, and natural language processing have been explored with the help of DL or deepened neural networks. Deep-RL in the URLLC is implemented by reducing the treatment time and improving the MLA chitecture by avoiding delays based on it. By adopting deep-RL for collaborative EMBB and URLLC scheduling, it was possible to improve high traffic demands and achieve QoS in 6LC, which shared collaboratively optimized bandwidth allocation and matching positions of URLLC users [45].

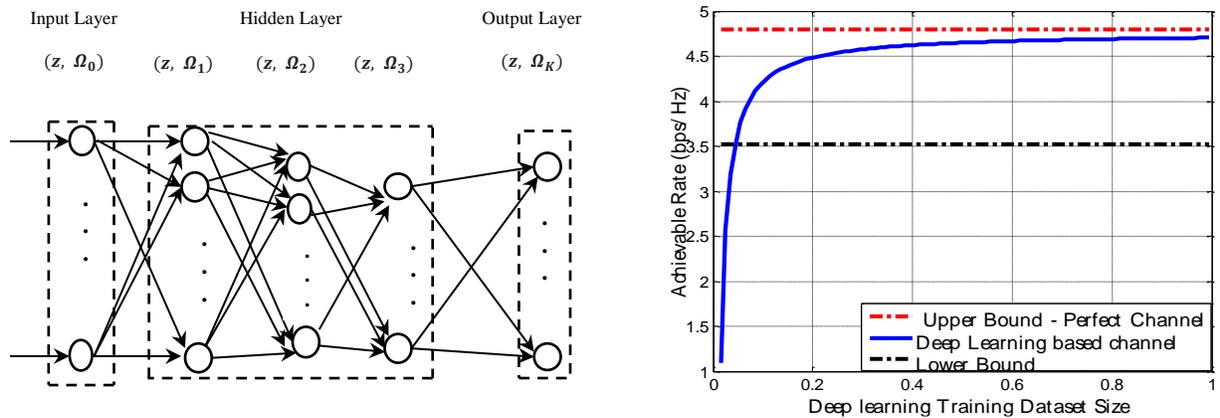

**FIGURE 4:** (a)Deep neural networks architectures for Link layer, (b) Attainable rate vs. the deep learning training dataset size.

## DL Platform for Mobile Networking with URLLC

We give the major primary underpinnings of DL network control in this section, and we argue that they are largely unexplored possibilities to explain mobile networking challenges. The use of a DL system with URLLC, which permits real-time connections on 6G networks, is required to successfully process massive data flowing from multiple sources. Based on user feedback and predictable outcomes, the URLLC improves network management [24]. Where the URLLC design focuses on enabling highly precise channel forecasts and high-speed traffic data that is essentially predictive, as well as directing the new deep learning development. The excellent prediction accuracy of DL with predictive URLLC enhances the consistency of the practical loss of the training data and the predictable loss based on the time-varying channels employed [7], [22], [30].

The major goal of this paper is to give a current picture of the URLLC, with a focus on the technical issues and solutions in 3GPP New Radio. We begin by describing the MLA for the DLLC, which handles demand-driven services and then discuss the operation of 6G networks for device exploration, as well as the transition to processing technologies.

## ENABLING DL IN MLA FOR URLLC

The URLLC is a brand-new application in 6G communication networks that will be vital to mission-critical IoT. In future 6G networks, where the facilitating mission-critical with severe requirements on end-to-end delay and reliability, achieving the data-driven for URLLC would be extremely difficult [46-51]. The data-driven DL may learn from a training data set and must achieve an appropriate range of wireless communication regulations [35]. Furthermore, the reliability requirement is determined based on a wide range of policies rather than the probability of packet loss ($10^{-5} \sim 10^{-7}$ )by utilizing a lengthy training period to achieve data−driven DL with extremely high dependability and low latency. However, obtain enough data−driven based on when a device's packet arrival rate is high by sending more than 107 packets over a lengthy period of time with a big number of training samples $10^7$ packets [47].

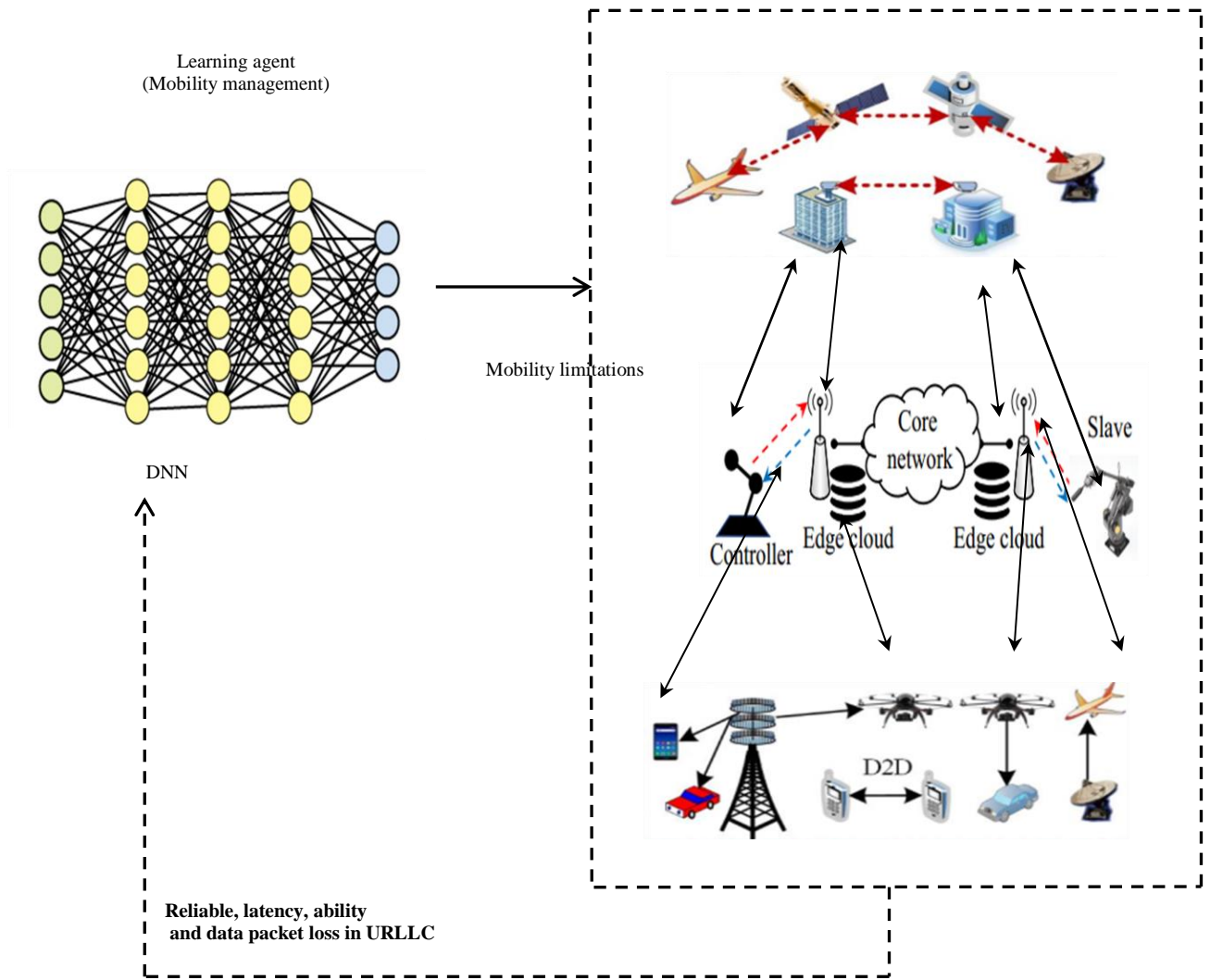

**FIGURE 5:** ML for permitting mobility supervision, scalability, and robustness in URLLC.

To solve this issue, we offer a MLA that supports data-driven edge and machine intelligence at the mobile user, cell, and network levels.

### Achieve of 6G Networks

The 6G networks are a data-driven network with close-knit, unrestricted wireless communication. Emerging 6G driving applications include a MLA that allows data-driven edge intelligence and device intelligence. The DL architecture improves the dependability and minimizes the latency of URLLC devices, as well as the deployment of all huge IoT devices involved. The data-driven DL has the potential to supply end-users with ultra-high data rates while also generating a suitable policy based on direct cost input from wireless networks [49]. To address these challenges in urgent need in 6G system as

### E2E Quality of Service

In 5G networks, the URLC reaches the level of latency and reliability of the third-generation partnership project (3gpp), which allows the latency to output sensitive data securely. In 6G networks, E2E latency and reliability requirements rarely improve, requiring network configuration to develop the best policies for short channel codes and short packets that allow DL in the URLLC. The quality of service should depend on productivity and quality [54]. The development of end-to-end technology for 6G can guarantee the quality of any physical experience and reach wireless everywhere and allow end-users to perceive and encircle a huge artificial brain that offers unlimited storage.

### Multi-Level Storage for Resources Allocation

Depending on the complete realization of IoT and DL in big data analytics in 6G networks, substantially more data can be transferred. Enabling AI for a future 6G network necessitates assessing the quantity of storage and computing capability, which is critical when using multi-level storage. The deployment of multi-level storage and computing resources in 6G enable scalable mobile-edge computing (MEC) [55]. Training data for deep neural networks deployed to mobile users in 6G based on the dense high-performance servers offered in the real-time system. To take advantage of the large amount of data available, DL has been used to learn from controlled big data, from large data analytics representations [56].

### Device Intelligence for User Level

Exploiting the number of URLLC is dependent on analyzing the reliability of device intelligence, local predicted information, and the ability to create decisions in mobile device anticipates such as tractability and transfer state. By examining the prediction error probability and restricted expected data in device intelligence at a user level, mobile users provide good decisions based on linked device activity on a network. The efficiency of user-level intelligence is determined by device intelligence technology's mobility management, which selects a prediction traffic state to enable end-users and then selects the best user association.

## Edge Intelligence for Cell Level

Artificial intelligence for edge intelligence is functionally required for the fast processing of massive amounts of data, and delivering a high level of AI to edge intelligence is dependent on a strong demand to combine edge computing and artificial intelligence. Deep-RL achieves improved package reliability and lower cost of industrial equipment in edge cell-level intelligence and optimizes the scheduler and value function at the access point using edge intelligence. It is recommended to improve the computational communication schedule by reducing the latency of the URLLC end and maintaining high AI to achieve high reliability for the URLLC at the cellular level. In addition, increasing level of data processing and capacity in the edge cell level depend on increasing levels of artificial intelligence and reducing times to a lower-level layer during training learning data.

## Cloud Intelligence at Network Level

Cloud intelligence-based computing allows mobile users to act as data collectors, sending data to cloud servers via an access point with regulated data preparation capabilities. Deep neural networks' artificial intelligence could improve channel estimates for large-scale channel gains by ensuring a high packet arrival rate from all user mobiles to the access point [57]. Because of the difficulty of predicting queue delays, DL used an offline data set in a central server algorithm based on the use of a mobility control entity to select the optimal user association in the revenue of a large-scale channel to a high packet arrival rate between mobile edge and each access point [11],[62-69]. Dynamic and self-learning systems in the data link layer and physical layer for the B5G and 6G physical layer are able to enhance performance and maximize user quality of the Service.

## Intelligence Sense Layer

A DL is utilized to supplement the stricter security and latency requirements. The most important tasks in 6G networks are the integration of sensing data in physical contexts and high capabilities with mobility. The sensing layer lies at the heart of the known routing protocols. For the physical network layer, the intelligence sensing layer is defined by independent intelligent sensing and information. Furthermore, increase the design flexibility of sensing rules and acknowledge that novel sensing functions require dedicated signal processing. Sensing is critical in cellular IoT, as highly precise sensing at IoT minimizes big data latency when transferring sense information through wireless communication networks.

# FUTURE RESEARCH DIRECTIONS

In this section, we offer AI data management based on application scripts based on the effective readability of the 6G network, as well as a MLA that handles the various bands supported in the previously discussed 6G.

**Estimation Efficiency:** Is a difficult problem for the DL to tackle, but one that can provide incrementally promising outcomes in wireless networking. Learning algorithms that achieve accurate identification are usually performed with improved computational complexity based on open inductive reasoning models. Learning algorithms are critical to efficiency because they work on data analysis in 6G and review transfer links with each IoT device designer. The DL for URLLC in the 6G wireless network has the best computing efficiency proposed.

**Hardware Communications:** Device creation is a significant challenge for 6G design, as radio access equipment and IoT devices will become a common way to meet future connectivity requirements. Allow 6G wireless networks to operate in the THz band, relying on less expensive hardware components, creating a low-cost policy with a good antenna and minimal operation, and making the device and algorithms invisible and multi-beam [18], [22]. Increase the processing time and frequency band it will increased the complexity of the hardware. To reduce the hardware complexity, the system, need to apply smart intelligence learning for action function and the state–action value function without requiring prior knowledge for the system.

**Terahertz communications:** Is a promising technology for 6G networks to support ultra-broadband. The THz band from 0.1 to 10 GHz was found as the gap between the microwave and optical spectra. To enable high-speed transmission of hundreds of Gbps for short-distance communication, increasing the capacity of the system requires increasing the bandwidth of the system to THz and increasing the spectrum efficiency by reducing the propagation loss [58-64]. In 6G, the amount of wireless data flow should be multiplied by several. To achieve the explosive growth of mobile data in 6G and establish reliable communication depends on apply THz, which achieve a terabit/second data rate, ultra-reliability and low latency without more spectral efficiency. The challenges necessitate spatial spectral efficiency and a broader radio frequency spectrum, both of which can be found exclusively in the THz and THz sub bands. To meet the needs for multi-Tb/s data rates, the combining of THz communications and sensing equipment is critical for future 6G cellular networks. To avoid significant path loss, the THz band will maintain a quiet frequency, mm-wave [49].

**Energy Management:** The biggest difficulty that 6G networks will face in the future is energy management, which would necessitate regulating the control and reduction of energy use. The energy management systems are meant to make the most of the energy that has been harvested. Energy efficiency is primarily a priority to reduce energy consumption by one bit (J / bit), because more information is required due to the intelligent connection to process large data and to operate a very large antenna. Advanced methods of energy management in 6G networks are very sharp. Artificial intelligence techniques have the potential to help these infrastructures and devices implement smart energy management strategies. The advantage of AI it is able to reduce hardware impairments and decrease power consumption in 6G by improving the scheduled channels or wait in the queuing list to arrive at this channel. Furthermore, neural networks employ AI and DL techniques to optimize energy management [21], [42], [63-69].

## CONCLUSIONS

In 6G networks, advanced energy management techniques are very sharp. Artificial intelligence approaches have the potential to assist these infrastructures and devices in implementing intelligent energy consumption control strategies. Furthermore, neural networks employ AI and DL techniques to optimize energy management. We then proposed some AI-enabled treatments for adopting various features of 6G network deployment and management, such as an MLA for URLLC in DL that enables data-driven edge intelligence and device intelligence at the mobile user, cell, and present AI technologies that improve 6G network performance based on its effective learning capabilities, such as compliant learning. THz communication is used in 6G ML to improve flawless channel quality estimation and excellent prediction accuracy for time-varying channels. Furthermore, the deployment of energy-harvesting systems for low-power consumption in 6G devices is dependent on future connectivity requirements.

## ACKNOWLEDGMENTS

This research is supported by Universiti Tun Hussein Onn Malaysia (UTHM) under the Multisciplinary Research (MDR) Grant votH243, and part by the Ministry of Higher Education Malaysia under the Fundamental Research Grant Scheme (FRGS/1/2019/TK04/UTHM/02/8) and RMC Fund (E15501) of Universiti Tun Hussein Onn Malaysia (UTHM).